\documentclass[twocolumn]{emulateapj}
\def\dnu{$\Delta\nu$}
\def\numax{$\nu_{\rm max}$}
\def\teff{$T_{\rm eff}$}
\def\logg{$\log g$}

\shorttitle{Asteroseismology of NGC~6791 and NGC~6819}
\shortauthors{Basu et al.}

\begin{document}

\title{Sounding open clusters: asteroseismic constraints from {\it Kepler} on the properties of NGC~6791 and NGC~6819}

\author{Sarbani Basu\altaffilmark{1}, Frank Grundahl\altaffilmark{2}, 
Dennis Stello\altaffilmark{3}, Thomas Kallinger\altaffilmark{4,5}, 
Saskia Hekker\altaffilmark{6}, 
Benoit Mosser \altaffilmark{7}, 
Rafael A. Garc{\'\i}a\altaffilmark{8}, 
Savita Mathur\altaffilmark{9}, 
Karsten Brogaard\altaffilmark{2}, 
Hans Bruntt\altaffilmark{2}, 
William J. Chaplin\altaffilmark{6},
Ning Gai\altaffilmark{10,1}, 
Yvonne Elsworth\altaffilmark{6}, 
Lisa Esch\altaffilmark{1}, 
Jerome Ballot\altaffilmark{11}, 
Timothy R. Bedding\altaffilmark{3}, 
Michael Gruberbauer\altaffilmark{12},
Daniel Huber\altaffilmark{3}, 
Andrea Miglio\altaffilmark{13}, 
Mutlu Y{\i}ld{\i}z\altaffilmark{14},
Hans Kjeldsen\altaffilmark{2}, 
J{\o}rgen Christensen-Dalsgaard\altaffilmark{2},
Ronald L. Gilliland\altaffilmark{15},
Michael M. Fanelli\altaffilmark{16},
Khadeejah A. Ibrahim\altaffilmark{17},
Jeffrey C. Smith\altaffilmark{18}
}

\altaffiltext{1}{Department of Astronomy, Yale University, P.O. Box
208101, New Haven, CT 06520-8101, USA}
\email{sarbani.basu@yale.edu}
\altaffiltext{2}{Department of Physics and Astronomy, Aarhus University, 8000 Aarhus C, Denmark}
\altaffiltext{3}{Sydney Institute for Astronomy (SIfA), School of Physics, University of Sydney, NSW 2006, Australia}
\altaffiltext{4}{Department of Physics and Astronomy, University of British Columbia, 6224 Agricultural Road, Vancouver, BC V6T 1Z1, Canada}
\altaffiltext{5}{Institute for Astronomy, University of Vienna, T\"urkenschanzstrasse 17, 1180 Vienna, Austria}
\altaffiltext{6}{School of Physics and Astronomy, University of Birmingham, Edgbaston, Birmingham B15 2TT, UK}
\altaffiltext{7}{LESIA, CNRS, Universit{\'e} Pierre et Marie Curie, Universit{\'e} Denis Diderot, 
Observatoire de Paris, 92195 Meudon, France}
\altaffiltext{8}{Laboratoire AIM, CEA/DSM-CNRS, Universit\'e Paris 7 Diderot, IRFU/SAp,
Centre de Saclay, 91191, Gif-sur-Yvette, France}
\altaffiltext{9}{High Altitude Observatory, National Center for Atmospheric Research, Boulder, CO 80307, USA}
\altaffiltext{10}{Department of Astronomy, Beijing Normal University, Beijing 100875, China}
\altaffiltext{11}{Laboratoire d'Astrophysique de Toulouse-Tarbes, Universit{\'e} de Toulouse, CNRS, 31400 Toulouse, France}
\altaffiltext{12}{Institute for Computational Astrophysics, Department of Astronomy and Physics, Saint Marys 
University, Halifax, NS B3H 3C3, Canada}
\altaffiltext{13}{Institut d'Astrophysique et G{\'e}ophysique de l'Universit{\'e} de Li{\`e}ge, All{\'e}e du 
six Ao{\^u}t, 17 B-4000 Li{\`e}ge, Belgium}
\altaffiltext{14}{Ege University, Department of Astronomy and Space Sciences, Bornova, 35100, \.Izmir, Turkey}
\altaffiltext{15}{Space Telescope Science Institute, 3700 San Martin Drive, Baltimore, MD 21218, USA}
\altaffiltext{16}{Bay Area Environmental Research Inst./NASA Ames Research Center, Moffett Field, CA 94035, USA}
\altaffiltext{17}{Orbital Sciences Corporation/NASA Ames Research Center, Moffett Field, CA 94035, USA}
\altaffiltext{18}{SETI Institute/NASA Ames Research Center, Moffett Field, CA 94035, USA}

\begin{abstract}

We present initial results on some of the properties of open
clusters NGC~6791 and NGC~6819 derived from asteroseismic data obtained
by NASA's {\it Kepler} mission. In addition to estimating the mass, radius and
\logg\ of stars on the red-giant branch of these clusters, we estimate the
distance to the clusters and their ages.
Our model-independent estimate of the  distance modulus of NGC~6791 is
$(m-M)_0= 13.11\pm 0.06$. We find $(m-M)_0= 11.85\pm 0.05$ for NGC~6819.
The average mass of stars on the red-giant branch of NGC~6791 is 
$1.20 \pm 0.01 M_\odot$, while that of NGC~6819 is $1.68\pm 0.03M_\odot$.
It should  be noted that we do not have data that cover the entire red-giant branch
and the actual mass will be somewhat lower.
We have determined model-dependent estimates of ages of these clusters.
We find ages between 6.8 and 8.6 Gyr for NGC~6791, however, most sets of models
give ages around 7Gyr. We obtain ages between 2 and 2.4 Gyr for NGC~6819.

\end{abstract}


\section{Introduction}
\label{sec:intro}

Oscillations in red giant stars in the field have been studied with Kepler
(Bedding et al. 2010, Huber et al. 2010, Kallinger et al. 2010b)
The first 34 days of science data from NASA's {\it Kepler} mission \citep{borucki10}
had shown that oscillations of red-giant stars in the clusters can also be detected \citep{stello10a}.
{\it Kepler} has been continuously observing stars in NGC~6819 and NGC~6791 for more than a year.
In this Letter we present results of  asteroseismic analyses of 
red-giant stars in NGC~6791 and NGC~6819 to
determine some basic properties of these clusters.  In particular, we 
derive model-independent estimates of the distance moduli of these clusters,
as well as the average mass of the stars on the red-giant branch. We also
derive estimates of ages of these stars.  We use
{\it Kepler} observations made between May 2009 and December 2009 (i.e Q1-Q3 data)
for this work.

The stars in these clusters, particularly those in NGC~6791, are faint and currently 
available {\it Kepler} data allow us to extract basic asteroseismic parameters
of only the bright red-giant branch and helium core burning red-clump stars. Although
red-giant stars have relatively large uncertainties in
asteroseismically determined properties, the uncertainties are still small enough
to make such an analysis viable. The fact that all cluster stars can be assumed
to have the same distance and age increases the precision of our results.

\section{Data and analysis}
\label{sec:data}

\begin{figure}
\epsscale{1.0}
\plotone{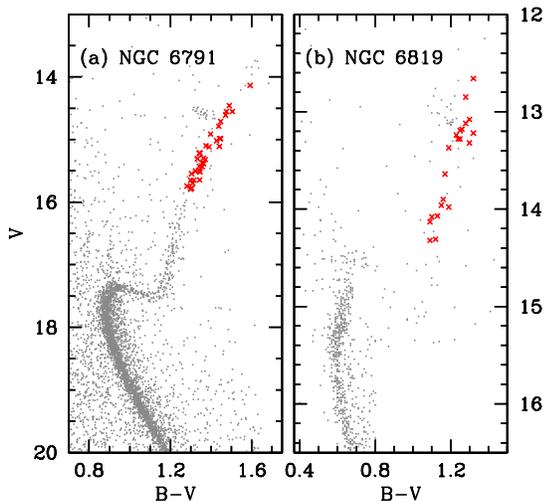}
\caption{The color-magnitude diagrams of NGC~6791 and NGC~6819 with the stars 
used in this work plotted in red. Photometry data for NGC~6791 are from Stetson et al.~(2003)
and for NGC~6819 from Hole et al.~(2009).}
\label{fig:cmd}
\end{figure}

We studied 34 stars in NGC~6791 and 21 stars in 
NGC~6819. Only confirmed members were selected (see Stello et al~2010a,b).
The stars studied in this work are shown in Figure~\ref{fig:cmd}. 
{ Note that although oscillation characteristics have been detected and
measured  in red-clump stars of both clusters,
we do not use them in this study for reasons explained later}.
Since
red-giant oscillations are slow, {\it Kepler} long-cadence observations
($\Delta t \simeq 30$ minutes; Jenkins et al. 2010) are adequate for our work.
The time series for each star was analyzed by five different pipelines:
SYD (Huber et al.~2009), COR (Mosser \& Appourchaux 2009), CAN (Kallinger et al. 2010a), A2Z (Mathur et al.~2010)
and OCT (Hekker et al. 2010).  A comparison of these pipelines can be found
in Hekker et al. (2011a).
We use two asteroseismic parameters in this study: the so-called large frequency separation, \dnu,
and the frequency of maximum power \numax.  Each pipeline produced estimates of these parameters.
For each cluster we adopted the results from the pipeline that
consistently gave the closest result to the average of the results obtained
by all pipelines. The uncertainties in \dnu\ and \numax\ returned 
by the pipeline were increased to reflect the variation
of \dnu\ and \numax\ values returned by the different pipelines.
We used those stars for which all pipelines returned a
result.  Details of the data preparation and analysis can be found in Hekker et al.~(2011b).

The large separation \dnu\ depends on the mean density of a star (Christensen-Dalsgaard 1988):
\begin{equation}
 \frac{\Delta\nu}{\Delta\nu_{\odot}}=\sqrt{\frac{M/M_{\odot}}{(R/R_{\odot})^{3}}}.
 \label{eq:delnu}
 \end{equation}
The scaling holds to within 3\% over most of the HR diagram (Stello et al.~2009a).
The frequency of maximum power in the oscillations power spectrum,
\numax, is related to the acoustic cut-off frequency of a star
(e.g., see Kjeldsen \& Bedding 1995; Bedding \& Kjeldsen 2003;
Chaplin et al. 2008) and scales as
 \begin{equation}
 {{\nu_{\rm max}}\over{\nu_{{\rm max},\odot}}}={{M/M_{\odot}}\over
 {(R/R_{\odot})^2\sqrt{(T_{\rm eff}/T_{{\rm eff},\odot})}}}.
 \label{eq:numax}
 \end{equation}
Data from CoRoT have verified this scaling  (Mosser et al.~2010).
Thus if \dnu, \numax\ and \teff\ are known, Equations~(\ref{eq:delnu}) and
(\ref{eq:numax}) represent two equations in two unknowns ($M$ and $R$)
and  can be solved to obtain  $M$, $R$ and  $g$.
However, Equations~(\ref{eq:delnu}) and (\ref{eq:numax}) are not constrained by the equations of stellar structure
and assume that all values of \teff\ are possible for a star of a given mass and radius resulting in
unrealistically large uncertainties in the derived mass and radius.
We  overcame this by using  so-called ``grid-based'' methods, where the characteristics
of stars are determined by searching among the models to get a ``best
fit'' for a given observed set of \dnu, \numax, \teff, and
[Fe/H].  Gai et al. (2011) have shown that  grid-based seismology
produces model-independent results for $R$ and \logg; there are small
biases in $M$ that can be minimized if metallicity is known, and that the
increased  precision is worth the small model-dependence in mass.

In this work we used three different implementations of grid-based methods --- the
Yale-Birmingham (YB) pipeline (Basu et al.~2010; Gai et al. 2011), 
a slightly modified version of the RADIUS pipeline of
Stello~(2009b) and the Seismic Fundamental Parameter (SFP) pipeline  of Kallinger et al.~(2010b).
YB was used with four different sets of models. One set, described by
 Gai et al.~(2011), was  based on YREC (Demarque et al. 2008).  The other sets were
those of Dotter et al.~(2008), Marigo et al.~(2008)
and models from the Yonsei-Yale (YY) isochrones (Demarque et al. 2004).
Additionally, a set of YREC models with $Y=0.30$ was also used to analyze
NGC~6791 data because of suggestions that the helium
abundance of NGC~6791 is high (Demarque et al. 1992; Brogaard et al. 2011). 
The RADIUS pipeline uses  models described by Stello et al.~(2009b) that were constructed
with the ASTEC code (Christensen-Dalsgaard et al.~2008).
The SFP pipeline is based on the BASTI models (Pietrinferni et al. 2004).
{ The different model
grids use different physics inputs. 
While the YREC and Dotter et al.
models include gravitational settling and diffusion of helium and heavy elements, 
YY models only include diffusion of helium.
The BASTI models, the Marigo et al. models  and the  Stello et al. models do not incorporate 
settling and diffusion of elements at all. Nuclear reaction rates and 
equations of state used to construct the models are different too.
The mixing length parameter for models in each grid was calibrated to a standard solar model constructed
using the corresponding stellar evolution code and it
differs somewhat from grid to grid because of differences in the physics used.
}
{ Equations~(\ref{eq:delnu}) and (\ref{eq:numax}) were used to calculate
\dnu\ and \numax\ for the models in all grids. }
Since the ASTEC and YY grids do not include core-helium burning red-clump stars,
we restricted our study to stars that are not likely to be in the core-helium
burning phase. 

Effective temperatures for stars in the two clusters were derived
using the color--temperature calibrations of Ramirez \& Melendez (2005).
All targets are sufficiently bright to have $JHK$ 
photometry from 2MASS (Skrutskie et al. 2006), allowing us to 
determine the temperatures based on the $(V-K)$ colors. We used $V$ estimates
of Stetson et al.~(2003) for NGC~6791 targets and Hole et al.~(2009)
for NGC~6819 targets.  Brogaard et al.~(2011) estimated the reddening towards 
NGC~6791 to be E$(B-V) = 0.16\pm0.02$
and that is the value adopted by us. 
We adopted a value of E$(B-V)=0.15$ (Bragaglia et al. 2001)
for NGC~6819. The uncertainty in each temperature estimate is
believed to be about 100K, and arises from uncertainties in photometry, reddening, as well as
uncertainty in the color--temperature relationship. 

We adopted a metallicity of
[Fe/H]$=+0.29$ for NGC~6791 (Brogaard et al.~2011)
and $+0.09$ for NGC~6819 (Bragaglia et al. 2001). We have assumed
uncertainties of $0.1$ dex, which is somewhat larger than the internal errors claimed
by each group.

\section{Results and Discussion}

For each cluster we first present estimates of stellar properties that are nearly model
independent, before looking at the question of age estimates which are
known to be model-dependent. As in the case of the basic seismic inputs \dnu\ and \numax, we
only use the results of one of the pipeline--grid combination  for all
parameters except age. The selected pipeline--grid combination was the one that
produced results that were consistently the closest to the average of the  results obtained from all pipelines
and grids of models. For each parameter of each star, the spread in results obtained from the
different pipelines and grids were added in quadrature to the formal uncertainty  obtained from the selected 
pipeline--grid combination. The basic data for both clusters are listed in Table~\ref{tab:res}.

\begin{table*} 
\caption{Properties of stars in NGC~6791 and NGC~6819. Stars are sorted according to \teff.
The assumed uncertainty in temperature is 100K.$^1$
}
\begin{center}
\vskip -0.25 true cm
{\scriptsize
\begin{tabular}{lccccccc}
\hline 
\noalign{\smallskip}
KIC ID & $T_{\rm eff}$ & $V$  & $\Delta\nu$& $\nu_{\rm max}$& Radius & Mass & $\log g$\\
      & (K)  & (mag)& ($\mu$Hz) & ($\mu$Hz) & ($R_\odot$) & ($M_\odot$) & (cgs) \\
\hline 
\multicolumn{8}{c}{NGC~6791}\\ 
\hline
 2437340 &4007 & 14.135 & $1.317\pm 0.054$ & $  8.028\pm0.257$ & $23.40^{+2.90}_{-2.86}$  &  $1.26^{+0.55}_{-0.53}$ &  $1.767^{+0.026}_{-0.017}$  \\
 2437444 &4186 & 14.553 & $2.459\pm 0.073$ & $ 18.658\pm0.496$ & $15.32^{+1.01}_{-0.37}$  &  $1.26^{+0.16}_{-0.11}$ &  $2.142^{+0.022}_{-0.023}$  \\
 2437816 &4215 & 14.459 & $2.348\pm 0.100$ & $ 16.948\pm0.484$ & $15.88^{+1.23}_{-0.61}$  &  $1.28^{+0.22}_{-0.15}$ &  $2.104^{+0.021}_{-0.025}$  \\
 2437507 &4246 & 14.554 & $2.568\pm 0.052$ & $ 20.654\pm0.282$ & $15.60^{+0.75}_{-0.77}$  &  $1.40^{+0.16}_{-0.14}$ &  $2.189^{+0.020}_{-0.020}$  \\
\hline 
\multicolumn{8}{c}{NGC~6819}\\
\hline
 5112880  &  4443 & 12.66 & $ 2.800\pm0.050$ & $ 26.650\pm1.065$ & $17.25^{+1.05}_{-1.04}$ & $2.04^{+0.35}_{-0.28}$ & $2.302^{+0.031}_{-0.031}$ \\
 4937576  &  4481 & 13.08 & $ 3.560\pm0.051$ & $ 32.290\pm1.516$ & $13.03^{+0.73}_{-0.76}$ & $1.66^{+0.16}_{-0.17}$ & $2.395^{+0.026}_{-0.026}$ \\
 5023732  &  4512 & 12.85 & $ 3.110\pm0.032$ & $ 27.450\pm1.234$ & $14.52^{+0.88}_{-0.85}$ & $1.75^{+0.24}_{-0.21}$ & $2.329^{+0.024}_{-0.023}$ \\
 5113041  &  4521 & 13.18 & $ 3.940\pm0.051$ & $ 37.570\pm1.384$ & $12.39^{+0.59}_{-0.51}$ & $1.65^{+0.14}_{-0.09}$ & $2.461^{+0.022}_{-0.022}$ \\
 5112734  &  4528 & 13.19 & $ 4.020\pm0.050$ & $ 40.210\pm1.545$ & $12.76^{+0.57}_{-0.62}$ & $1.74^{+0.20}_{-0.10}$ & $2.489^{+0.023}_{-0.024}$ \\
\hline
\end{tabular} 
}
\end{center}
$^1$ The complete table can be viewed online
\label{tab:res}
\end{table*} 

\subsection{NGC~6791}

\begin{figure*}
\epsscale{0.9}
\plotone{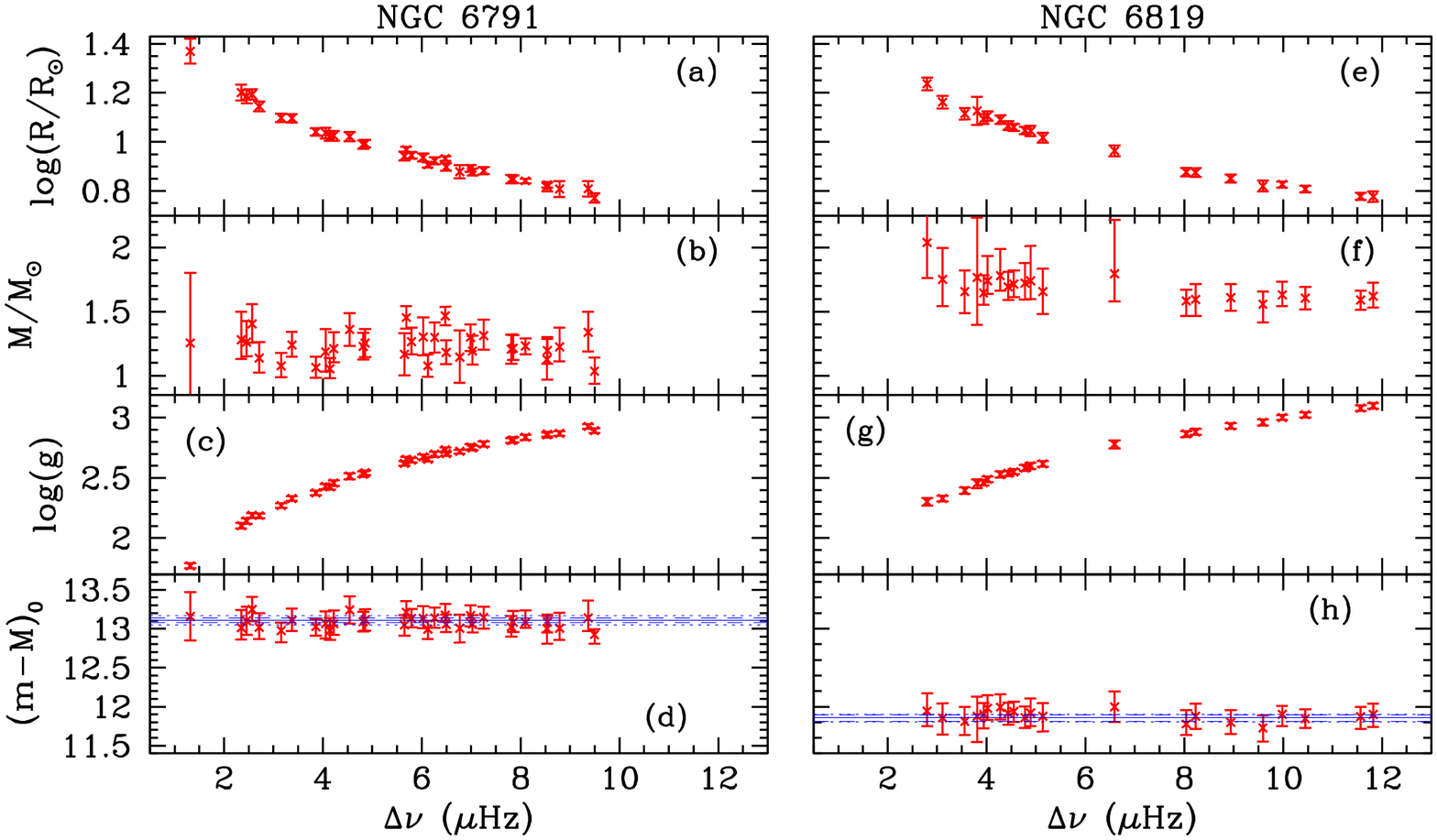}
\caption{The derived properties of stars in NGC~6791 (left
panels) and NGC~6819 (right panels) plotted against their large separations. We show the results for
 radius (panels a and e), mass (panels b and f), \logg\ (panels c and g) and
 extinction corrected distance modulus (panels d and h) for the individual stars. 
The distance modulus to the cluster has also been determined by adopting the prior
that all stars have to have the same distance modulus. The results are shown 
as the solid horizontal line. The dashed line shows the 1$\sigma$ limits of the
random error and the dotted lines show the result when the systematic error caused by
metallicity and reddening are added to the random error. { A simple average
of the DMs of individual stars gives a similar result, however,
the uncertainties in the mean are somewhat larger than the 1$\sigma$ limit shown by the dashed lines.}}
\label{fig:6791}
\end{figure*}

Results on temperature, radius, mass, \logg, and distance modulus (DM) for the stars in NGC~6791 are
shown in Figure~\ref{fig:6791}(a--d). The results are also listed in Table~\ref{tab:res}.
The results are derived with the adopted value of metallicity ([Fe/H]=$+0.29$) and
reddening (E$(B-V)=0.16$).

The stars in the sample cover a temperature range
of approximately 4000 to 4600K. The radii of the stars  increase almost monotonically
from  about $6R_\odot$ at the highest temperatures to about $24R_\odot$ at the
lowest temperatures.  This trend is completely consistent with what we expect from standard
red-giant models. { Although asteroseismic radius estimates of giants have  large
uncertainties (compared to other stars), we can still get radii to a precision of  about 5\% for all but six stars.}
Of the rest, five have precisions better to 8\% and one star has  a $\sim12$\% uncertainty. { The large
uncertainty in this case is a reflection of the spread in the results obtained by different pipelines;
we need more data to resolve this.}

Within uncertainties, all the stars have roughly equal masses.
The formal averaged-mass for the stretch of
the RGB branch for which we have data is $1.20 \pm 0.01 M_\odot$. The uncertainty on
the individual masses are larger, though generally less than 10\%. { Better estimates
of masses} will be possible once we model the detailed oscillation spectrum of
each star. { The data available so far does not yet allow us to do that.}

We obtain precise results for \logg\ --- with uncertainties less than
1\% for most stars.  Given that spectroscopic estimates of \logg\ for giants is uncertain, asteroseismology
gives us an alternative method of determining this quantity. Since \logg\ estimates are
required to determine stellar temperatures from spectra, this should enable us to obtain
better temperature estimates of these stars. 

NGC~6791 has been an object of repeated investigations ever since the first in-depth
study by Kinman (1965). There have been many estimates of the distance to this
cluster. Kinman (1965) found that
$(m-M)_0=13.55$ mag. Since then different studies have resulted in a wide
range of values for the DM that span the range of 12.6 mag (Anthony-Twarog \& Twarog 1985) to 13.6 mag 
(Harris \& Canterna 1981). Discussions of this can be found in Stetson et al.~(2003) and
Carraro et al.~(2006).
Our model-independent estimates of radius, along with temperature estimates, allow
us to determine the DM of the stars in the cluster. Better results are
obtained if, instead of calculating luminosity explicitly from radius and \teff,
we again do a grid search to determine luminosity, or the absolute visual magnitude
directly. Most of the grids include $M_V$ calculated from $L$ using a variety of
color-conversions. Where $M_V$ was not available in the grid, the luminosity
was converted to $M_V$ using the color-table of Lejeune et al.~(1997).

The DMs of our sample of NGC~6791 stars 
 are shown in  Figure~\ref{fig:6791}(e). Also shown is the
DM obtained assuming that all stars are at the same distance.
Using an extinction of $A_V=3.1$E$(B-V)$ we obtain $(m-M)_0= 13.11\pm 0.06$
for NGC~6791 assuming [Fe/H]=$+0.29$. There is a small dependence
of the DM on the adopted value of E$(B-V)$ and metallicity mainly through differences in temperature
estimates and that can be seen in Table~\ref{tab:dep}.
All estimates of the DM are very similar and the
adopted value of E$(B-V)$ does not change the DM drastically. Much of
the older literature on NGC~6791 finds/adopts E$(B-V)\simeq0.1$ which is
smaller than our adopted value of E$(B-V)=0.16$.

\subsection{NGC~6819}

Results for NGC~6819 are shown in Figure~\ref{fig:6791}(e--h) and also listed in Table~\ref{tab:res}.
The stars in our sample cover a temperature
range of 4450K to 4850K. The radii of the stars range from 6 to 17 $R_\odot$ and with
precisions to better than 5\% for all but 3 stars, and better than 6\% for all except one
star.
The average mass of these stars is $1.68\pm 0.03 M_\odot$. The uncertainty
on individual mass estimates is  less than 10\% in most cases.
The higher mass of these stars compared to those of NGC~6791 is indicative of a lower age
for this cluster. Again the uncertainty in \logg\ is less than 1\% in most cases.

The extinction-corrected DM for this cluster, when we assume [Fe/H]=$+0.09$ and
E$(B-V)=0.15$, is $(m-M)_0=11.85 \pm 0.05$. The DM, for the same value of metallicity
is 11.87 for E$(B-V)=0.10$ and 11.83 for E$(B-V)=0.20$. The E$(B-V)$ ranges were
selected from values used in literature about this cluster. If we assume  the same
reddening  but  [Fe/H]=$-0.10$ then
we obtain $(m-M)_0=11.91$.  Thus, as with NGC~6791, we can obtain precise values of
the DM even in the presence of some of the uncertainties in reddening. Changing metallicity
by larger than the assumed uncertainty does change the obtained value, but not by much.
The E$(B-V)$ and [Fe/H]-dependence can be seen in Table~\ref{tab:dep}. 

The DM values we have derived are somewhat smaller than those discussed in 
literature in the context of isochrone-fitting. Kalirai \& Tosi (2004) used
$(m-M)_0$ of 12--12.2 mag. Hole et al.~(2009) used $(m-M)_V=12.3$ mag, 
which for their adopted reddening of E$(B-V)=0.1$ mag is still somewhat larger
than what we find.

\begin{table}
\caption{Dependence of DM and age on E$(B-V)$ and [${\rm Fe/H}$]}
\begin{center}
\begin{tabular}{lccccc}
\hline
\noalign{\smallskip} 
\multicolumn{6}{c}{NGC~6791}\\
E$(B-V)$ & [Fe/H] & & $(m-M)_0$ & & Age (Gyr)\\
\noalign{\smallskip} 
\hline  
\noalign{\smallskip} 
0.10 & +0.29 & & $13.22\pm0.06$ & & $9.2^{+0.4}_{-0.5}$ \\
0.125& +0.29 & & $13.17\pm0.06$ & & $8.7^{+0.4}_{-0.5}$ \\
0.16 & +0.29 & & $13.11\pm0.06$ & & $8.0^{+0.5}_{-0.5}$ \\
0.18 & +0.29 & & $13.09\pm0.06$ & & $7.6^{+0.5}_{-0.5}$ \\
0.20 & +0.29 & & $13.04\pm0.06$ & & $7.2^{+0.5}_{-0.5}$ \\
0.25 & +0.29 & & $12.98\pm0.07$ & & $6.0^{+0.5}_{-0.6}$ \\
\noalign{\smallskip}
0.16 & +0.25 & & $13.13\pm0.06$ & & $8.1^{+0.5}_{-0.5}$ \\
0.16 & +0.29 & & $13.11\pm0.06$ & & $8.0^{+0.5}_{-0.5}$ \\
0.16 & +0.35 & & $13.07\pm0.06$ & & $7.9^{+0.5}_{-0.5}$ \\
0.16 & +0.40 & & $13.04\pm0.06$ & & $7.8^{+0.5}_{-0.5}$ \\
\noalign{\smallskip}
\hline  
\noalign{\smallskip} 
\noalign{\smallskip} 
\multicolumn{6}{c}{NGC~6819} \\
\noalign{\smallskip} 
E$(B-V)$ & [Fe/H] & & $(m-M)_0$ & & Age (Gyr)\\
\noalign{\smallskip}  
\hline   
\noalign{\smallskip}
0.10  & $ +0.10$& & $11.87\pm0.05$ & & $2.99^{+0.50}_{-0.25}$\\
0.125 & $ +0.10$& & $11.86\pm0.05$ & & $2.58^{+0.41}_{-0.21}$\\
0.15  & $ +0.10$& & $11.85\pm0.05$ & & $2.23^{+0.31}_{-0.17}$\\
0.175 & $ +0.10$& & $11.85\pm0.05$ & & $1.94^{+0.24}_{-0.14}$\\
0.20  & $ + 0.10$& & $11.83\pm0.06$ & & $1.70^{+0.18}_{-0.12}$\\
\noalign{\smallskip}
0.15  & $-0.15$& & $11.92\pm0.05$ & & $2.39^{+0.41}_{-0.17}$\\
0.15  & $-0.10$& & $11.91\pm0.04$ & & $2.36^{+0.39}_{-0.17}$\\
0.15  & $-0.05$& & $11.88\pm0.05$ & & $2.33^{+0.36}_{-0.17}$\\
0.15  & $ 0.00$& & $11.89\pm0.04$ & & $2.30^{+0.34}_{-0.17}$\\
0.15  & $+0.05$& & $11.86\pm0.05$ & & $2.27^{+0.33}_{-0.17}$\\
0.15  & $+0.10$& & $11.85\pm0.05$ & & $2.23^{+0.31}_{-0.17}$\\
0.15  & $+0.15$& & $11.84\pm0.05$ & & $2.18^{+0.30}_{-0.17}$\\
\noalign{\smallskip}   
\hline  
\end{tabular}  
\end{center} 
\label{tab:dep}
\end{table}

\subsection{Ages}

\begin{figure}
\epsscale{1.0}
\plotone{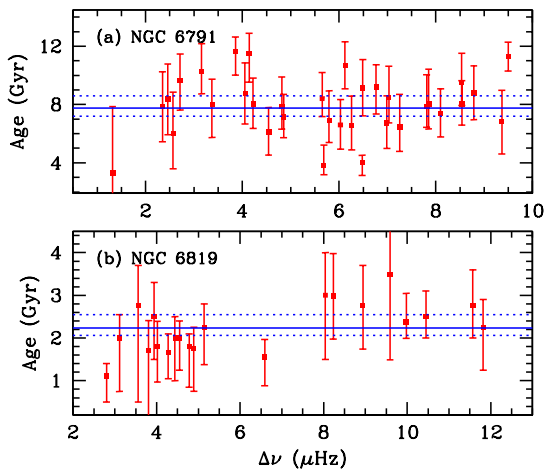}
\caption{Derived ages of stars in NGC~6791 (panel a) and NGC~6819 (panel b) using
YY models.  The points with
error-bars show the results for individual stars while the
horizontal lines in each panel is the result obtained applying the prior that
all stars in the cluster have the same age.}
\label{fig:age_both}
\end{figure}

The two easily observed asteroseismic quantities, \dnu\ and \numax, do
not have a direct dependence on age. Age estimates using these
two quantities thus rely on models.
For individual stars, the grid method gives large uncertainties,
around 25\% for most, but even larger for some. However,
it can be shown that applying an equal-age prior to  all stars in a given cluster
allows ages to be determined quite precisely (Gai et al. 2010).
Unlike the case of isochrone fitting, we can  derive the age of the
cluster without turn-off stars and the estimates are independent of distance. 
Dependence on reddening comes through
temperature calibrations.  Of the three pipelines used, at the moment only
YB allows the use of the prior of identical ages and hence only those
results are discussed. Figure~\ref{fig:age_both} shows the ages of the individual
stars in the two clusters as obtained with YY models. Also shown is the age
determined when the prior of identical age is applied. As is clear, the uncertainties
for individual stars is much larger than uncertainties for the cluster as a whole.

As expected from previous studies, our results show that NGC~6791 is considerably older
than NGC~6819. For the assumed metallicity of [Fe/H]=$+0.29$, we obtain fairly consistent
results from all the model grids used. The highest age obtained is $8.6\pm0.5$Gyr for YREC models,
though the YREC models with the high helium abundance gives $7.3^{+0.7}_{-0.4}$Gyr. We obtain
$8.0^{+0.8}_{-0.6}$Gyr using YY, and $7.2^{+0.8}_{-0.4}$ using  Dotter et al. models. 
Marigo et al.~(2008) models give a lower age ---  $6.8\pm0.4$Gyr. The difference between the
models lie in the helium abundance used as well as in the mixing length parameter used and
details of core overshoot. For any given grid of models,
changing the metallicity changes the estimated age, but by far the larger change occurs
when the assumed reddening is changed since that changes the estimated temperatures.
Table~\ref{tab:dep} lists the age estimates using YY models for different
values of E$(B-V)$ and [Fe/H].
Thus it is crucial to  estimate reddening precisely in order to derive a definitive age estimate for this
cluster. 

Our age estimates for NGC~6791 lie somewhat on the lower
side of most isochrone-fitting based estimates. Previous studies give a wide range of
ages between 8 and 12 Gyr (e.g. Carraro et al. 1994, 2006; Chaboyer et al. 1999, Carney et al.
2005, etc.). Among the more recent studies are those of  Carraro et al.~(2006), who concluded
that the age of this cluster should be in the range 7.5--8.5Gyr with preference given 
to the higher limit. Our estimated ages are, however, consistent with the age-estimate of 7Gyr
obtained by Brogaard et al~(2011).  This consistency is probably the  result of our adopting their
value of E$(B-V)$. Spectroscopic temperature estimates of stars in this cluster
are required to resolve this issue. Although our derived age is lower than most, it is not low enough to
alleviate the problem that this old metal-rich cluster poses for scenarios of Galactic chemical 
evolution.

Unlike the case of NGC~6791, our estimates for the age of NGC~6819 agree with those
determined through isochrone fitting. We derive ages  between 2.1 and 2.5 Gyr. Use of YREC models result in
an age estimate of $2.38^{+0.28}_{-0.20}$Gyr. YY gives $2.23^{+0.31}_{-0.17}$ Gyr. We found
$2.31^{+0.30}_{-0.16}$ Gyr with Dotter et al. models and $2.00^{+0.21}_{-0.14}$ Gyr with
Marigo et al. models. As with NGC~6791, the age estimates depend on the adopted value
of E$(B-V)$. The estimated age increases  if we decrease the assumed
E$(B-V)$. The estimated age is also sensitive to metallicity and
the dependence of age derived from YY models on E$(B-V)$ and [Fe/H] is listed in Table~\ref{tab:dep}.
Ages in literature vary from 1.8 to 2.5 Gyr (Kalirai \& Tosi 2004; 
Hole et al. 2009). 
We should be able to place much tighter constraints on the age of this cluster when we have data on
individual frequencies of subgiants stars in this cluster. The oscillation spectrum
of subgiant stars is extremely sensitive to age and modeling the spectra gives more
precise, though still somewhat model dependent,  age estimates (see e.g., Metcalfe et al. 2010).

\section{Conclusions}

Early seismic data from {\it Kepler} has allowed us to determine the 
basic properties of some of the red giants in NGC~6791 and NGC~6819, which
in turn has enabled us to determine the distance to those clusters in
a model independent manner. The final uncertainties are likely to improve
as we get data on more stars, thus  these results are just a { foretaste} of 
how we can use {\it Kepler} data to do high-precision studies of clusters.

Age estimates of the clusters are still somewhat uncertain, and the
adopted value of reddening plays a crucial role through its effect
on temperature estimates. We have, however, demonstrated that with only
a handful of stars it is possible to derive ages of clusters to levels of 
precision that rival ages obtained through isochrone fitting of the entire color-magnitude
diagram of a cluster. This holds promise for future studies of the two sparse clusters, NGC~6811
and NGC~6866, in the {\it Kepler} field.
For more reliable  age determinations,  we 
need better estimates of the effective temperature of cluster stars. It would
also help to have independent estimates of the mass of some stars in
binary systems, since that would constrain the search space.

The stars in these clusters are still being observed and we should soon have
time-series that are long enough for us to be able to extract individual frequencies.
Modelling the frequency spectrum of each  under the constraint that all stars need to be 
modeled with the same metallicity and that they should have the same age
will provide more stringent constraints on the properties of these clusters.

\acknowledgements
Funding for the {\it Kepler} mission is provided by NASA's Science Mission Directorate. 
The authors wish to thank the entire Kepler team, without whom these results 
would not be possible. We also thank all funding councils and agencies 
that have supported the activities of  Working Group 2 of the 
{\it Kepler} Asteroseismic Science Consortium.

{\it Facilities:} \facility{Kepler}

\clearpage
{\ }
\vskip 2 true in
\centerline{\LARGE MATERIAL FOR ONLINE TABLE}
\clearpage

\begin{table*} 
{ONLINE TABLE: Properties of stars in NGC~6791 and NGC~6819. Stars are sorted according to \teff.}
\begin{center}
\vskip -0.25 true cm
{\tiny
\begin{tabular}{lccccccc}
\hline 
\noalign{\smallskip}
KIC ID & $T_{\rm eff}^{(a)}$ & $V$  & $\Delta\nu$& $\nu_{\rm max}$& Radius & Mass & $\log g$\\
      & (K)  & (mag)& ($\mu$Hz) & ($\mu$Hz) & ($R_\odot$) & ($M_\odot$) & (cgs) \\
\hline 
\multicolumn{8}{c}{NGC~6791}\\ 
\hline
 2437340 &4007 & 14.135 & $1.317\pm 0.054$ & $  8.028\pm0.257$ & $23.40^{+2.90}_{-2.86}$  &  $1.26^{+0.55}_{-0.53}$ &  $1.767^{+0.026}_{-0.017}$  \\
 2437444 &4186 & 14.553 & $2.459\pm 0.073$ & $ 18.658\pm0.496$ & $15.32^{+1.01}_{-0.37}$  &  $1.26^{+0.16}_{-0.11}$ &  $2.142^{+0.022}_{-0.023}$  \\
 2437816 &4215 & 14.459 & $2.348\pm 0.100$ & $ 16.948\pm0.484$ & $15.88^{+1.23}_{-0.61}$  &  $1.28^{+0.22}_{-0.15}$ &  $2.104^{+0.021}_{-0.025}$  \\
 2437507 &4246 & 14.554 & $2.568\pm 0.052$ & $ 20.654\pm0.282$ & $15.60^{+0.75}_{-0.77}$  &  $1.40^{+0.16}_{-0.14}$ &  $2.189^{+0.020}_{-0.020}$  \\
 2569360 &4254 & 14.610 & $2.712\pm 0.050$ & $ 20.442\pm0.389$ & $13.95^{+0.67}_{-0.52}$  &  $1.14^{+0.12}_{-0.12}$ &  $2.187^{+0.018}_{-0.020}$  \\
 2436814 &4289 & 14.712 & $3.158\pm 0.046$ & $ 24.615\pm0.288$ & $12.55^{+0.51}_{-0.51}$  &  $1.08^{+0.10}_{-0.09}$ &  $2.270^{+0.018}_{-0.020}$  \\
 2436332 &4304 & 14.786 & $3.374\pm 0.054$ & $ 28.375\pm0.430$ & $12.49^{+0.50}_{-0.49}$  &  $1.24^{+0.10}_{-0.09}$ &  $2.329^{+0.021}_{-0.020}$  \\
 2436824 &4324 & 14.984 & $3.859\pm 0.059$ & $ 31.306\pm0.627$ & $10.98^{+0.40}_{-0.29}$  &  $1.06^{+0.09}_{-0.08}$ &  $2.376^{+0.020}_{-0.019}$  \\
 2436458 &4340 & 15.019 & $4.143\pm 0.049$ & $ 34.980\pm0.388$ & $10.47^{+0.36}_{-0.33}$  &  $1.05^{+0.08}_{-0.08}$ &  $2.425^{+0.020}_{-0.021}$  \\
 2435987 &4355 & 14.985 & $4.222\pm 0.050$ & $ 37.720\pm0.647$ & $10.60^{+0.50}_{-0.29}$  &  $1.21^{+0.13}_{-0.11}$ &  $2.456^{+0.023}_{-0.024}$  \\
 2436097 &4365 & 15.109 & $4.543\pm 0.045$ & $ 43.135\pm0.828$ & $10.50^{+0.47}_{-0.40}$  &  $1.36^{+0.13}_{-0.13}$ &  $2.515^{+0.025}_{-0.025}$  \\
 2436900 &4402 & 14.914 & $4.054\pm 0.059$ & $ 35.312\pm0.584$ & $10.87^{+0.55}_{-0.53}$  &  $1.19^{+0.18}_{-0.16}$ &  $2.428^{+0.021}_{-0.021}$  \\
 2437402 &4414 & 15.116 & $4.818\pm 0.057$ & $ 44.686\pm1.153$ & $ 9.76^{+0.39}_{-0.31}$  &  $1.23^{+0.10}_{-0.10}$ &  $2.531^{+0.022}_{-0.022}$  \\
 2437240 &4440 & 15.098 & $4.854\pm 0.038$ & $ 45.705\pm0.584$ & $ 9.83^{+0.35}_{-0.30}$  &  $1.26^{+0.11}_{-0.11}$ &  $2.543^{+0.020}_{-0.020}$  \\
 2436540 &4448 & 15.297 & $5.793\pm 0.062$ & $ 58.161\pm1.016$ & $ 8.81^{+0.28}_{-0.25}$  &  $1.27^{+0.11}_{-0.10}$ &  $2.648^{+0.019}_{-0.018}$  \\
 2438038 &4450 & 15.384 & $6.123\pm 0.048$ & $ 59.043\pm0.809$ & $ 8.06^{+0.21}_{-0.22}$  &  $1.08^{+0.09}_{-0.08}$ &  $2.655^{+0.018}_{-0.018}$  \\
 2437488 &4452 & 15.381 & $6.262\pm 0.057$ & $ 64.862\pm1.319$ & $ 8.38^{+0.29}_{-0.24}$  &  $1.30^{+0.12}_{-0.12}$ &  $2.697^{+0.022}_{-0.023}$  \\
 2437781 &4456 & 15.645 & $7.852\pm 0.068$ & $ 85.559\pm1.715$ & $ 7.05^{+0.26}_{-0.26}$  &  $1.21^{+0.10}_{-0.09}$ &  $2.816^{+0.022}_{-0.022}$  \\
 2437653 &4482 & 15.519 & $6.988\pm 0.062$ & $ 74.558\pm1.499$ & $ 7.79^{+0.24}_{-0.25}$  &  $1.30^{+0.10}_{-0.11}$ &  $2.757^{+0.022}_{-0.021}$  \\
 2570094 &4485 & 15.434 & $6.492\pm 0.066$ & $ 65.552\pm1.018$ & $ 7.91^{+0.28}_{-0.20}$  &  $1.18^{+0.10}_{-0.09}$ &  $2.703^{+0.017}_{-0.018}$  \\
 2436209 &4493 & 15.214 & $5.686\pm 0.038$ & $ 59.032\pm0.633$ & $ 9.30^{+0.25}_{-0.21}$  &  $1.45^{+0.09}_{-0.09}$ &  $2.656^{+0.020}_{-0.019}$  \\
 2569618 &4494 & 15.222 & $5.652\pm 0.038$ & $ 54.721\pm0.580$ & $ 8.75^{+0.37}_{-0.37}$  &  $1.17^{+0.16}_{-0.16}$ &  $2.620^{+0.020}_{-0.018}$  \\
 2570172 &4498 & 15.497 & $7.029\pm 0.070$ & $ 72.996\pm1.609$ & $ 7.55^{+0.26}_{-0.18}$  &  $1.19^{+0.12}_{-0.11}$ &  $2.750^{+0.020}_{-0.022}$  \\
 2437270 &4499 & 15.306 & $6.479\pm 0.039$ & $ 70.246\pm0.692$ & $ 8.56^{+0.19}_{-0.22}$  &  $1.47^{+0.07}_{-0.07}$ &  $2.734^{+0.020}_{-0.019}$  \\
 2438333 &4501 & 15.322 & $6.030\pm 0.051$ & $ 61.432\pm1.524$ & $ 8.64^{+0.33}_{-0.36}$  &  $1.30^{+0.15}_{-0.14}$ &  $2.674^{+0.019}_{-0.020}$  \\
 2437976 &4525 & 15.650 & $8.100\pm 0.043$ & $ 89.010\pm1.022$ & $ 6.94^{+0.15}_{-0.12}$  &  $1.23^{+0.06}_{-0.06}$ &  $2.837^{+0.021}_{-0.021}$  \\
 2436688 &4530 & 15.503 & $7.254\pm 0.064$ & $ 78.455\pm1.931$ & $ 7.64^{+0.27}_{-0.27}$  &  $1.31^{+0.13}_{-0.11}$ &  $2.781^{+0.023}_{-0.022}$  \\
 2437972 &4543 & 15.546 & $7.813\pm 0.067$ & $ 83.977\pm1.666$ & $ 7.05^{+0.28}_{-0.23}$  &  $1.21^{+0.12}_{-0.11}$ &  $2.811^{+0.021}_{-0.020}$  \\
 2438140 &4543 & 15.427 & $6.766\pm 0.063$ & $ 67.390\pm1.802$ & $ 7.56^{+0.47}_{-0.40}$  &  $1.15^{+0.21}_{-0.20}$ &  $2.721^{+0.014}_{-0.017}$  \\
 2436818 &4545 & 15.744 & $8.787\pm 0.335$ & $ 95.743\pm2.150$ & $ 6.43^{+0.51}_{-0.25}$  &  $1.22^{+0.15}_{-0.11}$ &  $2.869^{+0.021}_{-0.020}$  \\
 2437325 &4557 & 15.698 & $8.541\pm 0.049$ & $ 94.479\pm1.428$ & $ 6.69^{+0.18}_{-0.22}$  &  $1.20^{+0.11}_{-0.10}$ &  $2.864^{+0.034}_{-0.042}$  \\
 2570244 &4559 & 15.776 & $9.496\pm 0.169$ & $100.545\pm1.549$ & $ 5.92^{+0.26}_{-0.17}$  &  $1.03^{+0.11}_{-0.10}$ &  $2.894^{+0.019}_{-0.019}$  \\
 2437957 &4602 & 15.651 & $8.537\pm 0.044$ & $ 92.536\pm1.029$ & $ 6.56^{+0.29}_{-0.25}$  &  $1.13^{+0.16}_{-0.16}$ &  $2.856^{+0.021}_{-0.020}$  \\
 2437933 &4610 & 15.791 & $9.366\pm 0.305$ & $109.166\pm1.292$ & $ 6.44^{+0.47}_{-0.40}$  &  $1.34^{+0.16}_{-0.15}$ &  $2.929^{+0.020}_{-0.020}$  \\
\hline 
\multicolumn{8}{c}{NGC~6819}\\
\hline
 5112880  &  4443 & 12.66 & $ 2.800\pm0.050$ & $ 26.650\pm1.065$ & $17.25^{+1.05}_{-1.04}$ & $2.04^{+0.35}_{-0.28}$ & $2.302^{+0.031}_{-0.031}$ \\
 4937576  &  4481 & 13.08 & $ 3.560\pm0.051$ & $ 32.290\pm1.516$ & $13.03^{+0.73}_{-0.76}$ & $1.66^{+0.16}_{-0.17}$ & $2.395^{+0.026}_{-0.026}$ \\
 5023732  &  4512 & 12.85 & $ 3.110\pm0.032$ & $ 27.450\pm1.234$ & $14.52^{+0.88}_{-0.85}$ & $1.75^{+0.24}_{-0.21}$ & $2.329^{+0.024}_{-0.023}$ \\
 5113041  &  4521 & 13.18 & $ 3.940\pm0.051$ & $ 37.570\pm1.384$ & $12.39^{+0.59}_{-0.51}$ & $1.65^{+0.14}_{-0.09}$ & $2.461^{+0.022}_{-0.022}$ \\
 5112734  &  4528 & 13.19 & $ 4.020\pm0.050$ & $ 40.210\pm1.545$ & $12.76^{+0.57}_{-0.62}$ & $1.74^{+0.20}_{-0.10}$ & $2.489^{+0.023}_{-0.024}$ \\
 5024583  &  4540 & 13.12 & $ 3.810\pm0.041$ & $ 38.150\pm9.710$ & $13.38^{+1.91}_{-2.40}$ & $1.77^{+0.47}_{-0.37}$ & $2.455^{+0.034}_{-0.041}$ \\
 5112744  &  4546 & 13.28 & $ 4.440\pm0.048$ & $ 44.740\pm1.490$ & $11.65^{+0.49}_{-0.48}$ & $1.70^{+0.12}_{-0.10}$ & $2.537^{+0.020}_{-0.021}$ \\
 5112948  &  4554 & 13.22 & $ 4.280\pm0.050$ & $ 43.960\pm1.415$ & $12.32^{+0.52}_{-0.49}$ & $1.78^{+0.21}_{-0.12}$ & $2.529^{+0.024}_{-0.024}$ \\
 5024297  &  4582 & 13.28 & $ 4.550\pm0.061$ & $ 46.030\pm0.832$ & $11.47^{+0.39}_{-0.39}$ & $1.72^{+0.11}_{-0.10}$ & $2.552^{+0.021}_{-0.021}$ \\
 5024404  &  4614 & 13.24 & $ 4.780\pm0.040$ & $ 49.260\pm1.405$ & $11.15^{+0.40}_{-0.39}$ & $1.72^{+0.16}_{-0.13}$ & $2.583^{+0.024}_{-0.024}$ \\
 5023931  &  4618 & 13.32 & $ 4.890\pm0.050$ & $ 51.270\pm2.118$ & $11.08^{+0.55}_{-0.49}$ & $1.74^{+0.27}_{-0.14}$ & $2.601^{+0.025}_{-0.027}$ \\
 5111940  &  4647 & 13.37 & $ 5.140\pm0.061$ & $ 52.890\pm1.401$ & $10.40^{+0.49}_{-0.48}$ & $1.66^{+0.18}_{-0.17}$ & $2.616^{+0.025}_{-0.024}$ \\
 5024405  &  4668 & 13.98 & $ 8.230\pm0.080$ & $ 97.840\pm2.565$ & $ 7.51^{+0.31}_{-0.29}$ & $1.60^{+0.12}_{-0.13}$ & $2.882^{+0.025}_{-0.025}$ \\
 5024312  &  4707 & 13.90 & $ 8.040\pm0.081$ & $ 92.990\pm3.404$ & $ 7.53^{+0.31}_{-0.33}$ & $1.58^{+0.09}_{-0.12}$ & $2.865^{+0.024}_{-0.025}$ \\
 5024512  &  4727 & 13.64 & $ 6.590\pm0.061$ & $ 76.370\pm2.905$ & $ 9.20^{+0.46}_{-0.45}$ & $1.80^{+0.42}_{-0.22}$ & $2.779^{+0.033}_{-0.033}$ \\
 5024143  &  4743 & 14.07 & $ 9.590\pm0.100$ & $115.760\pm5.644$ & $ 6.61^{+0.34}_{-0.36}$ & $1.56^{+0.10}_{-0.14}$ & $2.962^{+0.026}_{-0.026}$ \\
 5023845  &  4761 & 13.96 & $ 8.940\pm0.091$ & $107.910\pm3.655$ & $ 7.10^{+0.27}_{-0.30}$ & $1.61^{+0.11}_{-0.10}$ & $2.932^{+0.024}_{-0.024}$ \\
 5113441  &  4788 & 14.31 & $11.570\pm0.116$ & $152.580\pm3.674$ & $ 6.00^{+0.20}_{-0.19}$ & $1.59^{+0.07}_{-0.08}$ & $3.081^{+0.023}_{-0.023}$ \\
 5112072  &  4814 & 14.08 & $ 9.980\pm0.081$ & $126.510\pm2.904$ & $ 6.71^{+0.20}_{-0.20}$ & $1.63^{+0.10}_{-0.09}$ & $3.003^{+0.020}_{-0.022}$ \\
 5111718  &  4816 & 14.13 & $10.450\pm0.100$ & $132.960\pm2.452$ & $ 6.43^{+0.19}_{-0.18}$ & $1.61^{+0.09}_{-0.09}$ & $3.024^{+0.020}_{-0.021}$ \\
 5024240  &  4843 & 14.32 & $11.820\pm0.143$ & $158.070\pm6.521$ & $ 5.98^{+0.30}_{-0.26}$ & $1.62^{+0.11}_{-0.09}$ & $3.100^{+0.024}_{-0.023}$ \\
\hline
\end{tabular} 
}
\end{center}
{$^{(a)}$ Assumed uncertainty of 100K \hfill\break}
\label{tab:online}
\end{table*}

\end{document}